\documentclass[a4paper,11pt]{article}
\pdfoutput=1 

\usepackage{jcappub} 

\usepackage[T1]{fontenc} 
\DeclareMathAlphabet{\mathbit}{OT1}{cmr}{bx}{it}

\usepackage{units}
\usepackage{siunitx}
\usepackage{todonotes}
\usepackage{subfig}
\usepackage[export]{adjustbox}
\usepackage{placeins}
\usepackage{multirow}
\usepackage{booktabs}
\usepackage[version=4]{mhchem}
\usepackage{algpseudocode}

\title{CNNs for enhanced background discrimination in DSNB searches in large-scale water-Gd detectors}

\author[a,1]{D. Maksimovi\'c,\note{Corresponding author.}}
\author[a]{M. Nieslony,}
\author[a]{M. Wurm}

\affiliation[a]{Institute of Physics and Excellence Cluster PRISMA$^{+}$, Johannes Gutenberg-Universität Mainz,\\ 55099 Mainz, Germany}

\emailAdd{damaksim@uni-mainz.de}
\emailAdd{mnieslon@uni-mainz.de}
\emailAdd{wurmm@uni-mainz.de}

\abstract{Gadolinium-loading of large water Cherenkov detectors is a prime method for the detection of the Diffuse Supernova Neutrino Background (DSNB). While the enhanced neutron tagging capability greatly reduces single-event backgrounds, correlated events mimicking the IBD coincidence signature remain a potentially harmful background. Neutral-Current (NC) interactions of atmospheric neutrinos potentially dominate the DSNB signal especially in the low-energy range of the observation window that reaches from about 12 to 30\,MeV.

The present paper investigates a novel method for the discrimination of this background. Convolutional Neural Networks (CNNs) offer the possibility for a direct analysis and classification of the PMT hit patterns of the prompt events. Based on the events generated in a simplified SuperKamiokande-like detector setup, we find that a trained CNN can maintain a signal efficiency of 96\% while reducing the residual NC background to 2\% of the original rate. Comparing to recent predictions of the DSNB signal and measurements of the NC background levels in Super-Kamiokande, the corresponding signal-to-background ratio is about 4:1, providing excellent conditions for a DSNB discovery.}

\makeatletter
\gdef\@fpheader{}
\makeatother

\begin{document}
\maketitle
\flushbottom

\section{Introduction}
\label{sec:intro}

The first detection of the Diffuse Supernova Neutrino Background (DSNB) will be a mile stone for low-energy neutrino astronomy. While large-scale scintillator detectors like the upcoming JUNO or future Theia experiments are promising DNSB observatories \cite{An:2015jdp,Sawatzki:2020mpb}, the first glimpse of the signal can be expected from the current Phase VI of the Super-Kamiokande (SK) experiment \cite{Marti-Magro:2020igk}. By doping the water target with gadolinium, SK-Gd obtains an enhanced neutron capture signal permitting to isolate the fast coincidence signals induced by inverse beta decays (IBDs) \cite{Beacom:2003nk}. This new feature can be used to suppress most of the single-event backgrounds that have before limited the sensitivity of the DSNB search in SK \cite{Malek:2002ns,Bays:2011si,Zhang:2013tua}.

Despite the greatly improved conditions, DSNB detection is still beset by important backgrounds. IBDs induced by reactor and atmospheric $\bar\nu_e$'s define a narrow detection window from roughly 12\,MeV to 30\,MeV. Within this observation window, the most prominent background is posed by Neutral-Current (NC) interactions of atmospheric neutrinos on oxygen in the water that are able to mimic the IBD coincidence signature. Based on recent SK results for the expected NC reaction rates \cite{Wan:2019xnl}, it seems likely that this background may have a significant impact on DSNB detection in the visible energy range from 12 to 20\,MeV. 
 
Recent activities in the frame of the ANNIE experiment at Fermilab have shown that Convolutional Neural Networks (CNNs) are an excellent tool for event discrimination in water Cherenkov detectors \cite{msc:david:2020,Back:2017kfo}. In the present paper, we apply these findings to the DSNB signal and atmospheric NC background events expected for a SK-Gd-like simulation setup. We demonstrate that the discrimination power of a MC-trained CNN is sufficient to isolate the DSNB signal with high efficiency, achieving a signal-to-background ratio far greater than 1. Since the CNN does not rely on ''conventional'' reconstruction and discrimination parameters as the event vertex or Cherenkov angle \cite{Bays:2011si}, its application may offer a systematically independent tool for background discrimination.

The paper is organized as follows: We first describe the expected DSNB signal and the general observation window in a water-Gd Cherenkov detector (Sec.~\ref{sec:dsnb}). We next discuss our model of the atmospheric neutrino NC background that we link to recent results for atmospheric-neutrino NC event rates observed in SK (Sec.~\ref{sec:atmo_nc}). We go on to describe the detector simulation realized via the open-source software framework WCSim (Sec.~\ref{sec:simulation}) that we used to generate signal and background data. Section \ref{sec:cnn} describes the architecture, training and validation of the CNN used for data classification. Finally, we present our result for the suppression of the atmospheric neutrino NC background (Sec.~\ref{sec:results}) before concluding in section \ref{sec:conclusions}.

\section{DSNB signal}
\label{sec:dsnb}

While not yet observed, the DSNB signal can be predicted based on our current understanding of core-collapse Supernovae (SNe) and cosmology. Following the calculations layed down in Ref.~\cite{Beacom:2010}, the expected differential electron antineutrino ($\bar{\nu}_e$) flux can be expressed by the integral
\begin{align} \label{eq:diffflux}
    \frac{{\rm d}\phi}{{\rm d}E_{{\bar{\nu}}_e}}\left(E_{\bar{\nu}_e}\right)=\ \frac{c}{H_{0}}\int_{0}^{z_{\rm max}} {R_{\rm SN}\left(z\right)\varphi\left(E_{{\bar{\nu}}_e}\left(1+z\right)\right)\frac{{\rm d}z}{\sqrt{{\Omega}_{\Lambda}+\Omega_{\rm M}(1+z)^3}}}. 
\end{align} 
In the calculation, $R_{\rm SN}\left(z\right)$ is the red-shift dependent SN rate that can be integrated up to an upper limit of $z_{\rm max}$ and $\Omega_\Lambda$, $\Omega_{\rm M}$ and $H_0$ are cosmological parameters. To first order, it can be assumed that the neutrino energy spectrum $\varphi(E_{\bar{\nu}_e})$ can be represented by a Fermi-Dirac statistic with an average temperate $T$ for the $\bar{\nu}_e$ component.

To obtain the expected prompt positron signal spectrum ${\rm d}N/{\rm d}E_{e^+}$, we regard the product of the differential flux (\ref{eq:diffflux}), the IBD cross section $\sigma_{\rm IBD}\left(E_{{\bar{\nu}}_e}\right)$ (adopted from \cite{Beacom:2010}) and -- in the case of SK-Gd -- the number of free protons $N_p\approx1.5\cdot10^{33}$ in a 22.5-kt fiducial volume:
\begin{align} \label{eq:difflux_pos}
    \frac{{\rm d}N}{{\rm d}E_{e^+}}=\frac{{\rm d}\phi}{{\rm d}E_{{\bar{\nu}}_e}}\left(E_{\bar{\nu}_e}\right) \sigma_{\rm IBD}\left(E_{{\bar{\nu}}_e}\right) N_p.
\end{align}

\begin{figure}[!h]
\centering
\includegraphics[width=0.8\textwidth]{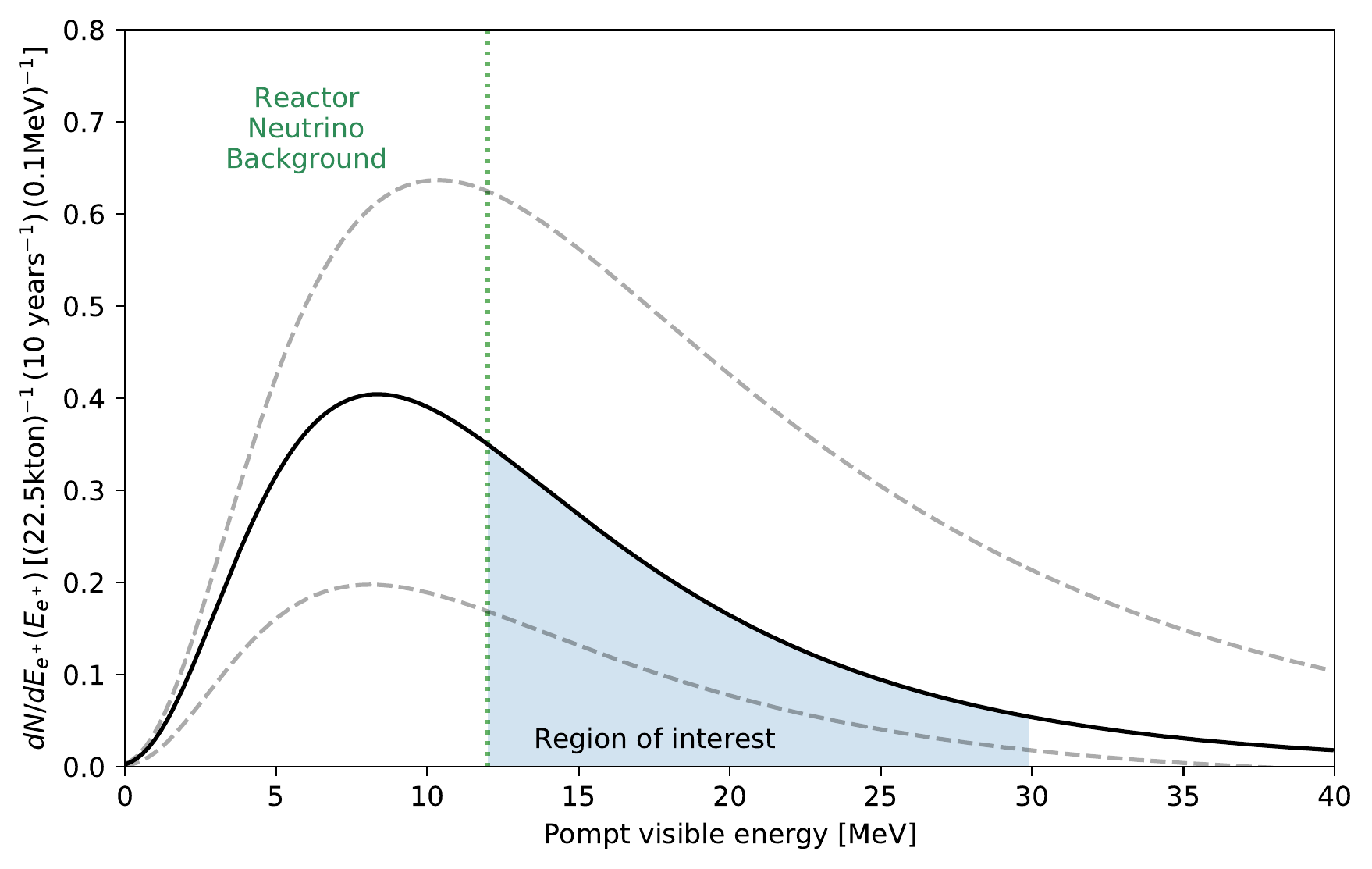}
\caption{Visible energy spectrum of the DSNB as observed in the IBD channel. We use the estimate for the DSNB flux estimated by Kresse {\it et al.} \cite{Kresse_2021}. We show the fiducial model ({\it solid black line}) along with upper and lower bounds for the prediction ({\it grey dashed lines}). In SK-Gd, the DSNB signal is dominated by the reactor neutrino background at energies below 12\, MeV \cite{Beacom:2003nk}. In the region of interest (12-30\,MeV), the fiducial model predicts 20 events in 10 years for a fiducial volume of 22.5 kt and a delayed neutron detection efficiency of {67\,\%}.}
\label{fig:DSNB_positrons}
\end{figure}
From the variety of available DSNB models (e.g.~\cite{Ando:2004hc,Beacom_2006,Beacom_2007}), we use here the relatively recent calculation from Ref.~\cite{Kresse_2021}. Kresse {\it et al.} predict the $\bar\nu_e$ spectrum taking into account individual SN neutrino spectra for a wide range of progenitor stars with initial masses between 9 and 120 solar masses. Moreover, they include the effects of black hole formation in failed explosions. In the following, we adopt the ''fiducial model'' to provide a standard DSNB neutrino rate and spectrum against which we will compare the atmospheric-neutrino NC background. To give an impression of the potential variability of the DSNB signal, Fig.\ref{fig:DSNB_positrons} shows not only the event spectrum expected for the fiducial model but as well spectra  that mark lower and upper signal expectations as indicated in Ref.~\cite{Kresse_2021}. 

In an SK-Gd like detector, DSNB detection is limited to the energy region between 12 and 30 MeV. At the lower energy end, IBDs induced by reactor $\bar\nu_e$'s produce an irreducible background. At higher energies, the $\bar\nu_e$ component of the atmospheric neutrino flux also induces IBD reactions and starts dominating the DSNB signals from about 25--35\,MeV. For the fiducial DSNB model, we find that 2 IBD events are expected per year within this observation window, while 0.9 and 4.9 events per year are expected for the lower and upper bounds of the flux prediction, respectively. The quoted rates already include a neutron detection efficiency of 67\% for the delayed neutron tag after all relevant cuts, corresponding to 0.1\% of gadolinium loading \cite{Kunxian:2015ymr}. This number may vary by a few percent depending on the final Gd mass fraction and the inclusion of above-threshold neutron captures on hydrogen in the data set. 

Requiring a single delayed signal from neutron capture on Gd is the key ingredient to suppress most of the backgrounds that will otherwise cover the DSNB signal in the observation window. Still, even including the neutron tag, the window is not free of background. The two major sources of background remaining are atmospheric neutrino NC interactions in the range up to 20\,MeV and so-called ''invisible muons'' in the energy range above. While estimates for the NC background level and its  discrimination by CNNs are the main theme of this paper, we shortly comment here on the latter background: ''invisible muons'' denote CC interactions of atmospheric neutrinos with low-energy final-state muons below the Cherenkov threshold. Those reactions can mimic the IBD signature with the prompt event originating from the Michel decay electron and the delayed event resulting from the neutrino-induced conversion of e.g.~a free proton. A corresponding spectral prediction is shown in Fig.~\ref{fig:class_power}. Since we do not expect that a CNN can pick up on the minute difference between the Cherenkov signals of a positron (IBD) and a Michel electron, we do not include this background component in our discrimination analysis.

\section{Atmospheric NC background}
\label{sec:atmo_nc}

As stated above, NC interactions of atmospheric neutrinos constitute an important background for the detection of the DSNB signal. These reactions can mimic the Inverse Beta Decay (IBD) coincidence signature by emitting one or several gammas and one or more neutrons in nuclear breakup reactions. While the hadronic fragments of an NC interactions are mostly below Cherenkov threshold, the gammas produced in de-excitation processes of the remnant nucleus are identified as the prompt event within the expected visible energy window. Neutrons are ejected from the nucleus in the primary neutrino interaction and thus provide the necessary delayed tag of the coincidence signature. 

 The Neutral Current Quasi-Elastic (NCQE) scattering process is the most prominent among the relevant NC atmospheric interactions on oxygen and was most recently characterized by the T2K experiment \cite{Abe_2019}. In this case, a nucleon is knocked out of the nucleus and in turn creates a proton or neutron hole in the oxygen nucleus. Different de-excitation processes can take place depending on the initial shell of the ejected nucleon.  In this work, we rely on the atmospheric event generator of the \texttt{GENIE} neutrino simulation framework (version 3.0.6)~to obtain the NCQE neutrino reactions and subsequent de-excitation channels \cite{Andreopoulos:2009rq,Andreopoulos:2015wxa}. The underlying atmospheric neutrino spectra in the energy range from 0.1 to 10\,GeV are taken from \texttt{FLUKA} \cite{Battistoni:2002ew}. Within \texttt{GENIE}, the oxygen nuclear de-excitation model is based on the theoretical calculations by H.~Ejiri~\cite{Ejiri:1993} and measurements of de-excitation gamma branching ratios by the $^{16}\mathrm{O} (p,p) ^{15}\mathrm{N}$ scattering experiment RCNP-E148 that were conducted in 2005~\cite{Kobayashi:2005ut}. 

Table \ref{tab:nucshells} provides an overview of the gamma rays emitted in the de-excitation of the final state nucleus. The branching fractions describing the probabilities for a nucleon to be ejected from a given shell are based on the theoretical calculations of H.~Ejiri. 
It is worth to note that measurements by a $^{16}\mathrm{O} (e,e'p) ^{15}\mathrm{N}$ scattering experiment from 1994 indicate that the spectroscopic factors that are expressed by the branching fractions might be $\sim$30\% lower than the theory predictions \cite{Leuschner:1994}. However, this effect will be compensated by the data-driven total normalization we apply based on the measured de-excitation energy spectrum (Sec.~\ref{sec:simulation}).


\begin{table}
\caption{Overview of the gamma rays resulting from the de-excitation of final state nuclei in NCQE interactions. The table lists branching ratios and typical energies as a function of the ejected nucleon and its initial shell in the \ce{^{16}O} nucleus.}\label{tab:nucshells}
\centering
\begin{tabular}{l|ccc}
\toprule
 & & \multicolumn{2}{c}{gamma energy [MeV]}\\
 O shell    &  branching & proton ejected & neutron ejected\\
\midrule
$1p_{1/2}$ & 25\% & no gamma & no gamma\\ 
$1p_{3/2}$ & 48\% & 6.32\,MeV (87.2\%) & 6.18\,MeV (91.7\%) \\
& & 9.93\,MeV (6.4\%) & no gamma (8.3\%) \\
& & no gamma (6.4\%) &  \\
$1s_{1/2}$ & 25\% & (3.09 - 7.34\,MeV) (32\%) & 7.03\,MeV (8\%)\\
& & no gamma (68\%) & no gamma (92\%)\\
\bottomrule
\end{tabular}
\end{table}

While no de-excitation is expected in case a nucleon is knocked out of the ($1p_{1/2}$) shell, a hole in the ($1p_{3/2}$) shell will lead to the emission of gamma rays. Typical energies range to $\sim$10\,MeV in case of proton knock-out, while a neutron knock-out from this shell would be followed by a 6.18\,MeV gamma in almost all cases.
A nucleon hole in the ($s_{1/2}$) state leaves the remnant nucleus in an energy level above the particle emission threshold, which means that de-excitation can happen through a variety of channels, including gammas, nucleons and alpha particles. In the case of proton holes, the gamma emission probabilities are extracted from the measurements of the RCNP-E148 experiment that specifically investigates such proton holes in the ($s_{1/2}$) shell of oxygen. Since there are no equivalent measurements for neutron hole processes in oxygen, the theoretical calculations of Ejiri are used to predict both the available de-excitation energy levels and emission probabilities in this case. More recent results for the RCNP-E148 experiment are available and include a few more de-excitation channels and slightly more precise results by means of a larger data sample \cite{Kobayashi:2006gb}. However, the differences in the emission probabilities and measured energy levels are very minor and deemed negligible in the context of this study.

Besides the NCQE reaction, another important NC background channel for the DSNB search are NC-$n\pi$ reactions, in which at least one pion is produced in the primary neutrino interaction. The pion is subsequently absorbed by an oxygen nucleus, which is put into an excited state and can in turn emit gammas. Such reactions are estimated to constitute $\sim$\,34\,\% of the overall NC background in the relevant energy window of the DSNB search \cite{Wan:2019xnl}. Due to their very similar final-state topologies, we expect only minor differences in the performance of the CNNs for the NC-n$\pi$ events in comparison to NCQE events. We therefore do not simulate such reactions explicitly, but evaluate their impact by scaling the overall NC background rate as described in Sec.\ \ref{sec:simulation}.

\section{Simulation framework}
\label{sec:simulation}

The simulations of the DSNB signal and atmospheric NCQE background event samples are performed in \texttt{WCSim}, an open-source \texttt{Geant4}-based framework for modeling particle interactions and the corresponding detector response in Water Cherenkov detectors \cite{WCSim:2020,Agostinelli:2002hh}. The \texttt{Geant4} version in use is \texttt{Geant4.10.1}, with the \texttt{FTFP\_BERT\_HP} physics list to model secondary interactions of neutrons including neutron captures and inelastic scattering processes. 
A cylindrical detector geometry with Super-Kamiokande dimensions ($r\sim17$\,m, $h\sim36$\,m, $\sim11150$ PMTs \cite{Fukuda:2002uc}) is chosen for the instrumented water tank. The detection medium is gadolinium-loaded water with a Gd mass fraction of 0.1\%. Water transparency values, Quantum Efficiency curves for the PMTs and PMT dark noise rates are assumed to be the same as in the original Super-Kamiokande simulation \texttt{SKDETSIM}. In this context, the detector response of the Super-Kamiokande geometry in \texttt{WCSim} has been validated to reproduce the behavior of the \texttt{SKDETSIM} simulation \cite{validation_wcsim}. The trigger threshold is configured to be 25 PMT hits within a 200\,ns window, which roughly corresponds to a lower energy threshold of 3-4\,MeV~for triggered events. Both the events of the DSNB signal and atmospheric NCQE sample are distributed uniformly across a Fiducial Volume of dimensions ($r\sim15$\,m, $h\sim32$\,m). This volume provides a minimum distance of 2 meters to any detector wall, corresponds to a fiducial mass of 22.5\,kt and is chosen in accordance to reference \cite{Bays:2011si}. In the final experiment, events inside this fiducial volume could be selected based on the well-probed vertex reconstruction algorithms available for SK and other water Cherenkov detectors. 

\paragraph{DSNB signal.} Inverse beta decays induced by the DSNB are simulated in \texttt{WCSim} by placing the two final state particles ($e^+$ and $n$) of the Inverse Beta Decay reaction at the same vertex in the cylindrical water tank and sampling uniform positions within the volume. The positron energy is modeled according to the expected DSNB antineutrino energy distribution that was described in section \ref{sec:dsnb}. Apart from the energy threshold, kinematics are neglected. 

\paragraph{NCQE background.} The atmospheric event generator of \texttt{GENIE} (v3.0.6) is used to simulate the primary interactions of the atmospheric neutrinos with the nuclei in the water tank according to the \texttt{FLUKA} atmospheric neutrino spectra, as described in section~\ref{sec:atmo_nc}. The \texttt{GENIE} event generation step provides information about which primary particles are produced in the interaction alongside their respective kinematic properties, already including nuclear final state interactions and potential de-excitation processes in the nucleus. All primary particles that are generated in the neutrino interaction are propagated to \texttt{WCSim} to model further interactions of the particles in the water and the detector response. 

In addition to the gammas originating from the de-excitation of the remnant nucleus in the primary neutrino interaction, the prompt signal includes as well secondary gammas that are produced in inelastic neutron and proton reactions with neighboring oxygen nuclei. Depending on the incident energy, these secondary interactions can produce additional neutrons from the break-up of oxygen nuclei \cite{Ashida:2019nhd}.

\paragraph{Rate normalization.} The expected rate for the DSNB signal has been derived in section~\ref{sec:dsnb}. For the atmospheric NC sample, we adjust the rate normalization to recent measurements of the NC atmospheric neutrino spectrum by Super-Kamiokande \cite{Wan:2019xnl}. The referenced measured spectrum includes atmospheric NC events with at least one detected neutron, split into NCQE, NC-$n\pi$, and accidental background contributions. We apply the energy-dependent efficiency correction provided in Ref.~\cite{Wan:2019xnl} to obtain the expected number of NCQE atmospheric neutrino events that produce at least one neutron. By selecting an equivalent data sample with at least one neutron out of our simulated NCQE events, we can carry out a data-driven normalization of our background sample. We use the same  normalization method for the NC-$n\pi$ sample, obtaining in this case as well the spectral shape from Ref.~\cite{Wan:2019xnl}.

\paragraph{Event selection.} For the prompt events, we apply an energy window from 12\,MeV to 30\,MeV delimited by other IBD backgrounds (sec.~\ref{sec:dsnb}). Energies for both signal and background samples are expressed in terms of visible energy, a value proportional to the number of detected photoelectrons and representative of the equivalent positron energy for a given charge. While DSNB signals are caused by a single positron, the signals of NCQE reactions are usually produced by several lower-energy gammas that add up to a visible energy in the selection range. Multiple gammas often result from secondary nuclear interactions of knock-out nucleons. For the delayed events, we do not reconstruct neutron captures on gadolinium explicitly but assume a uniform detection efficiency of $\varepsilon = 67\%$ after all relevant cuts \cite{Kunxian:2015ymr}. Since IBD events produce only a single neutron, this efficiency scales the rate linearly. The situation is somewhat more complex in case of the NCQE and NC-$n\pi$  events that often produce more than one neutron. In this case, we consider the probability that out of $n$ neutrons exactly one is detected ($n=1$), and scale the residual event rate accordingly.


\begin{figure}[b]
\centering
\includegraphics[scale=0.7]{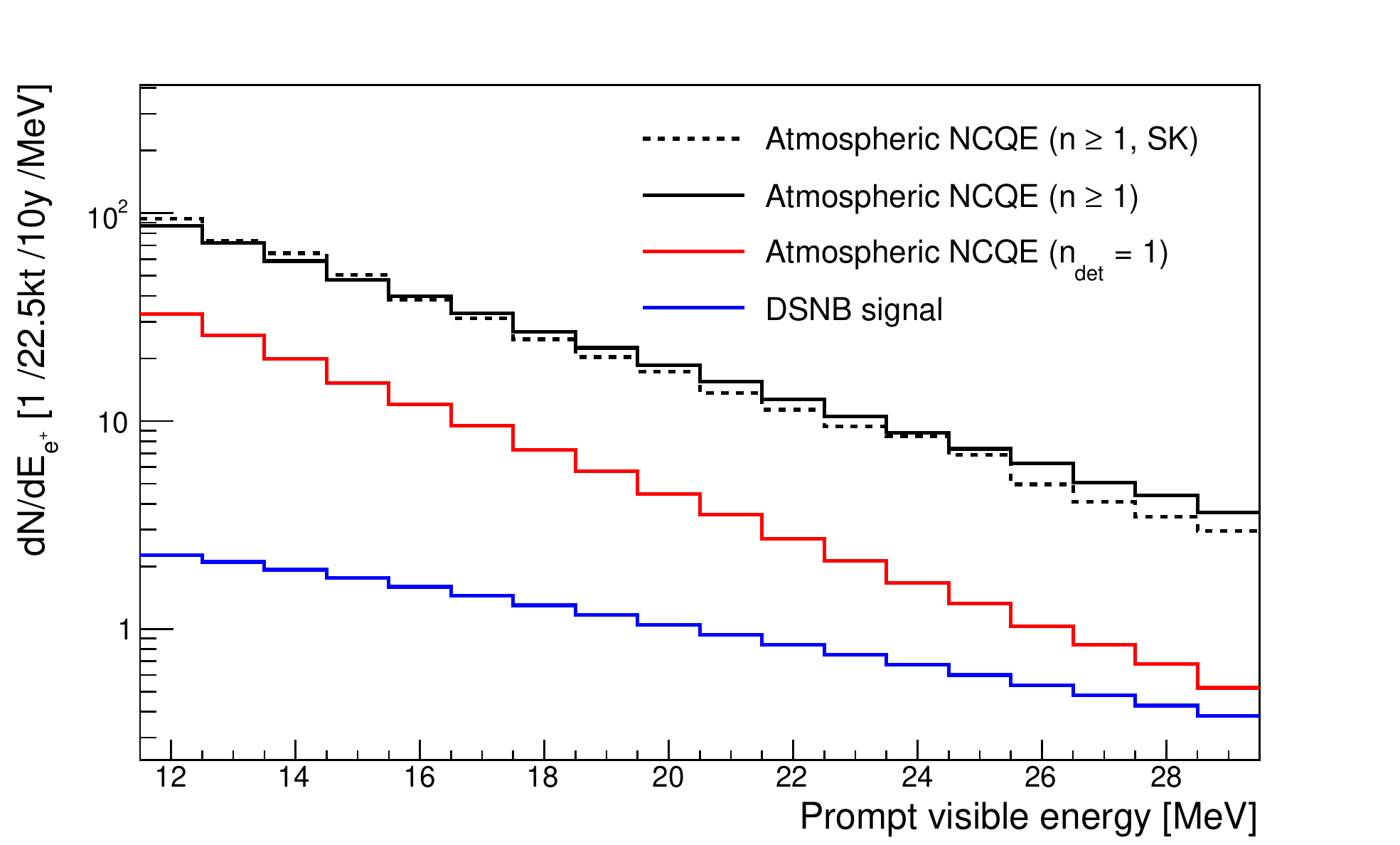}
\caption{The expected event rates for the simulated DSNB signal ({\it blue solid line}) and the atmospheric NCQE background samples. While the selection cuts for one detected neutron ({\it red line}) are used for the final background data sample, a different selection cut of at least one neutron ({\it black solid line}) is used to normalize the spectra based on the measurements by the Super-Kamiokande experiment ({\it black dashed line}) \cite{Wan:2019xnl}.}
\label{fig:dsnb_ncqe_normalization}
\end{figure}

\paragraph{Event spectra.} The resulting visible-energy spectra for DSNB and atmospheric background events are shown in figure~\ref{fig:dsnb_ncqe_normalization}. The shape of the simulated NCQE spectrum with $n\geq1$ largely reproduces the shape of the SK reference spectrum. The background rate exceeds the DSNB signal expectation by a factor of $\sim$24. Selecting only events with $n=1$ decreases the background rate significantly. Note that the spectrum softens as a result. The ratio of NCQE to DSNB events corresponds to about 7:1, increasing to 11:1 when including NC-$n\pi$ events. Particularly the lower end of the energy window is dominated by background, while the signal and background contributions reach similar levels for visible energies approaching $30$\,MeV. Note as well that the exact shape of the NC background spectra will feature a slight dependence on the value of the detection efficiency $\varepsilon\approx 0.67$ since it scales the relative contributions of event spectra with more than one final-state neutron to the effective spectrum with exactly $n=1$ detected. However, the corresponding impact on CNN classification efficiency is expected to be low (in accordance with the results displayed in Fig.~\ref{fig:class_neutron}).

 While we assume that general event cleaning cuts for removing other backgrounds such as spallation events and Michel electrons from the real data sample will still be necessary to reduce the data set to DSNB and atmospheric event candidates, we do not apply any further reconstruction or analysis cuts (e.g.~on the Cherenkov angle) beyond the one-neutron requirement. This explains why the simulated NC background spectrum looks different in terms of its spectral shape and event number from other other references (such as \cite{MuonRate}) that include such NCQE-specific selection cuts. The CNN-based background-reduction method presented in the next sections is such complementary to more ''traditional'' analysis approaches for distinguishing between DSNB and atmospheric event topologies.

\section{Event classification by CNN}
\label{sec:cnn}
By now, machine learning techniques and especially neural networks are widely used tools for event reconstruction and classification in particle physics experiments. Convolutional Neural Networks (CNNs) are especially well-adapted to these tasks since they are optimized to recognize recurrent features in stacks of digital images. Most neutrino detectors and in particular SK represent their light detection patterns by color-scale maps of PMT hits in 2D projections of the detector surface. These event displays are an ideal input for a CNN-based classification of signal and background events based on the differences in their respective event topologies. 

In the present context of the DSNB search in a water-Gd detector, we show here a promising approach to handle the NCQE atmospheric background through supervised learning. We first describe the preparation of the data as input for the CNN (Sec.~\ref{sec:data_prep}) before we turn to the network architecture and training (Sec.~\ref{sec:cnn_arch}). After a first glance at the network performance (Sec.~\ref{sec:cnn_performance}), we study possible biases in the CNN performance (Sec.~\ref{sec:bias}). The results for the classification of DSNB and NCQE data are described in Sec.~\ref{sec:results}.

\subsection{Data preparation}
\label{sec:data_prep}
For classification by the CNN, the DSNB and NCQE data generated as described in Sec.~\ref{sec:simulation} is converted into a pixelized image we call a ''hit map'', conceptually close to a conventional event display. The collected charge (photo electrons, p.e.) per PMT and first photon arrival time are translated into two hit maps in which each PMT is assorted to a specific pixel. Pixels in both hit maps are normalized to a range from 0 to 1. In the case of charge information, normalization is done relative to the largest p.e.~number detected in the present event. This prevents the CNN picking up on the absolute scale of the observed charge signal and thus largely prevents an energy bias in the classification (Sec.~\ref{sec:bias}).
Arrival times are scaled to values from 0 to 1 within a pre-defined time window of 270 ns surrounding the event trigger time with a 20 ns long pre-trigger window and a 250 ns long post-trigger-window. Early trials showed that the fixed time interval is essential to prevent the CNN from a classification based on the early presence of dark noise hits in one event category.

The topology of the detector is mapped onto the 2D image plane by a projection of the cylindrical tank surface. Like for traditional event displays, the cylinder mantle is cut open and rolled out horizontally over the range $\varphi=0..2\pi$. The top and bottom PMTs are rearranged by spreading them along the $\varphi$ axis in such a way that the original circular arrangement of the PMTs around the coordinate poles is broken down into four smaller cut-outs. This choice effectively reduces the height of the hit map required for representing the bottom and top lid. 
While this projection creates apparent discontinuities (e.g.~at the coordinate transition for $\varphi=2\pi,0$) that sometimes split and spatially separate topological features like a Cerenkov cone, the performance of the CNN is not diminished by these effects since these features of the hit maps are recognized and connected during training. 

\begin{figure}[!t]
\centering
\includegraphics[width=\textwidth,trim={0 2.29cm 0 0cm},clip]{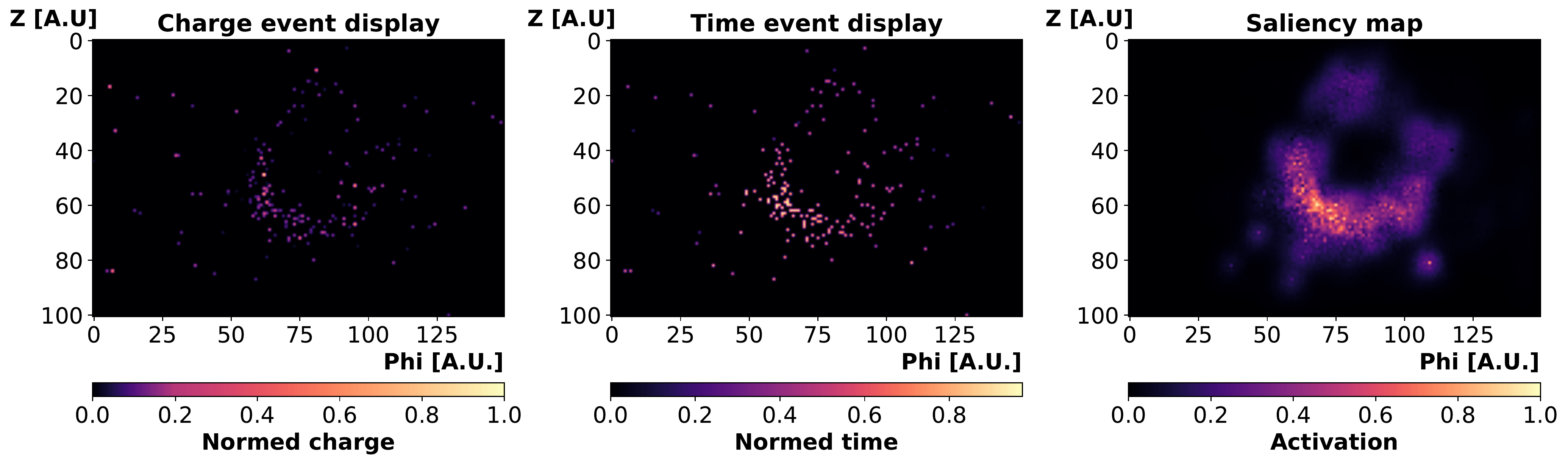}
\includegraphics[width=\textwidth,trim={0 0cm 0cm 0cm}]{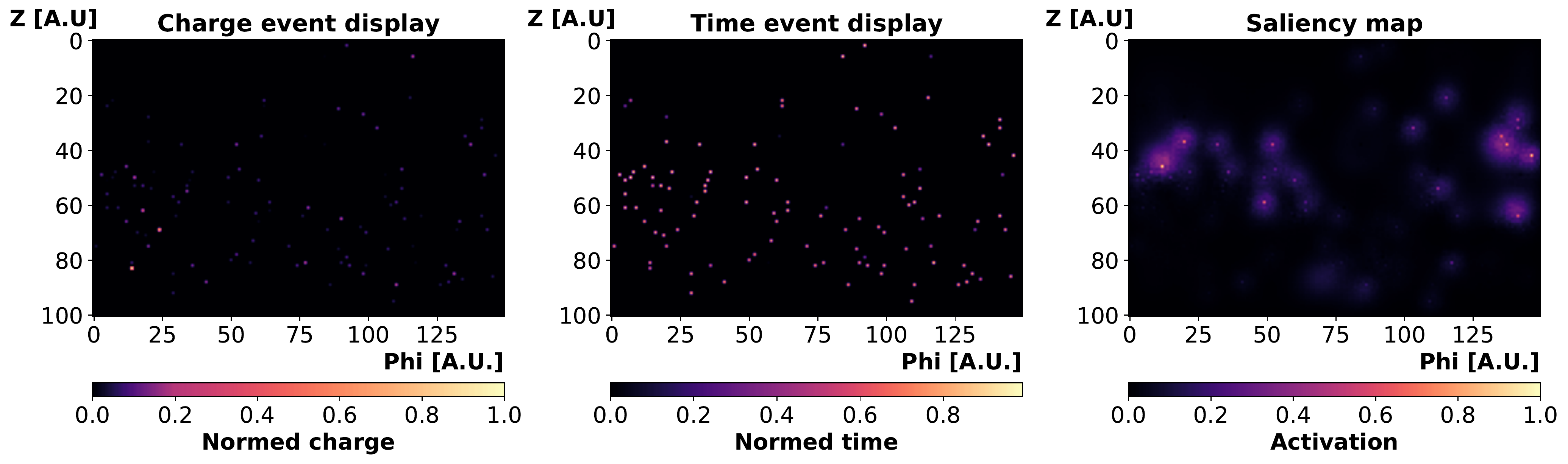}
\caption{A representative pair of event hit maps prepared for classification by the CNN. The upper panels show a DSNB event (22\,MeV) with a single Cherenkov ring, while the lower panels display a NCQE event with several weak rings from multiple gammas and a total visible energy of 19\,MeV. From left to right, normalized charge and time hit maps and saliency maps are shown. The latter highlight the regions of the maps that the CNN uses for classification.}
\label{fig:event_displays}
\end{figure}

Figure \ref{fig:event_displays} displays the corresponding hit maps for event charge and timing for two exemplary events of DSNB and NCQE categories. To the human eye, the difference between a single strong Cherenkov ring from an IBD positron and several weak rings created by several low-energy gammas in the NCQE event is immediately apparent. In anticipation of the later sections, the two right-most panels show so-called saliency maps that visualize the emphasis given by the trained CNN to specific regions of the 2D projections in the classification of these events. While the attention of the network is centered on the Cerenkov ring in case of the DSNB events, it is quite spread out for the NCQE map.



\subsection{Network architecture and training}
\label{sec:cnn_arch}

\begin{figure}[!h]
\centering
\includegraphics[scale=0.5]{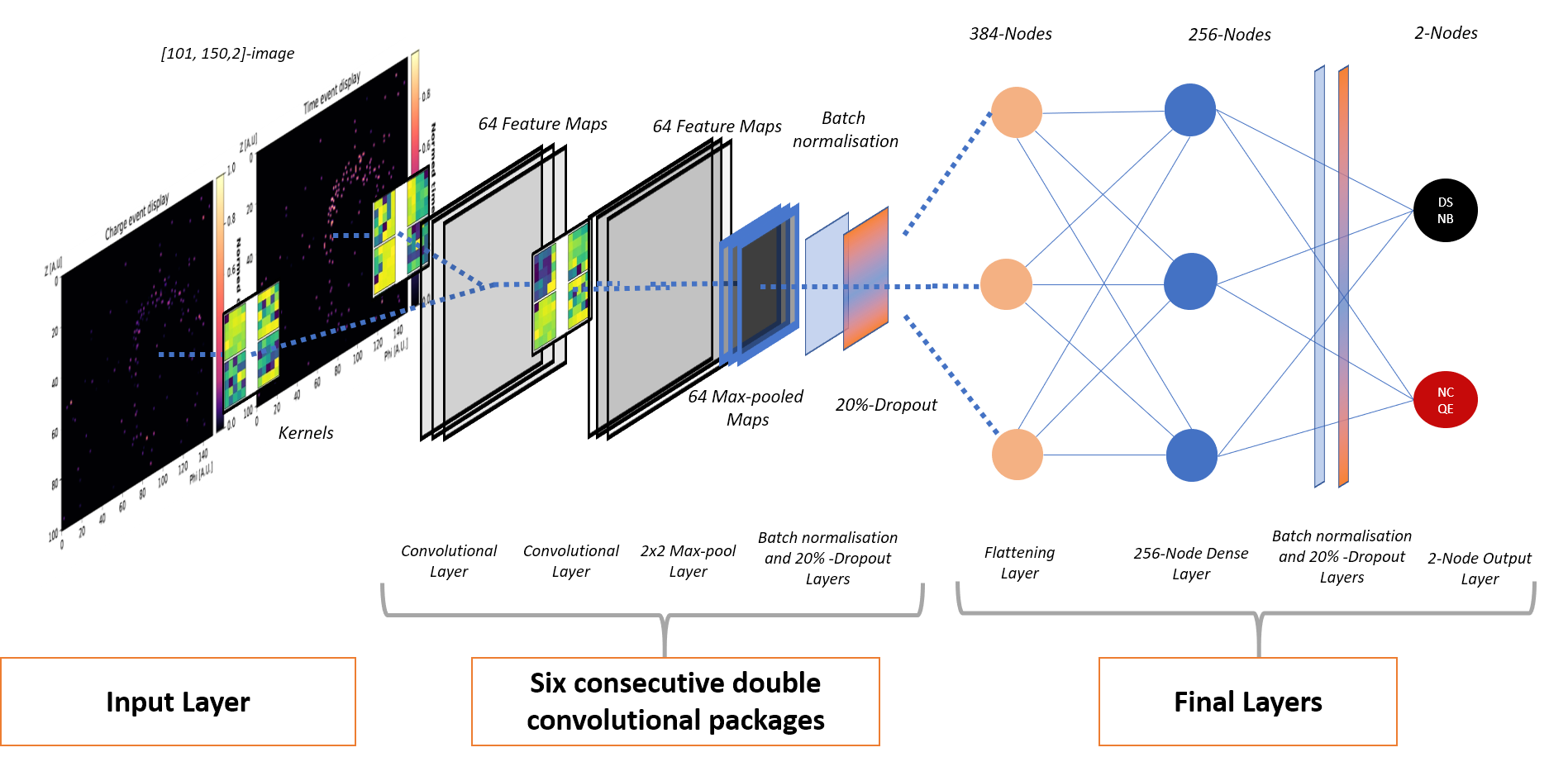}
\caption{Schematic model of the CNN architecture. The first convolutional layer of the first package receives the two separate time and charge hit maps and utilizes 64 kernels for each channel. The resulting feature maps are combined and form a common (101,150,64)-image to be analyzed by the second convolutional layer.}
\label{fig:NetworkArc}
\end{figure}

\paragraph{CNN architecture.} To build and train the CNN, we used Python 3.7 \cite{10.5555/1593511} with the Keras library \cite{Keras} and the Tensorflow framework \cite{tensorflow-whitepaper}. A schematic view of the network is given in Figure \ref{fig:NetworkArc}. It features a total of 726k trainable and 1.2k non-trainable parameters and consists of six consecutive "64-double convolutional packages". Each package consists of two successive convolutional layers with 64 kernels each, followed by a 2-dimensional (2,2)-Maxpooling layer, a BatchNormalisation layer and a final Dropout layer in four of the six packages which is set to 20 \%. The purpose of convolutional layers is to detect features within their input image. A convolutional layer applies kernels/filters to small successive sections of their input map and stores the results in a so-called feature map. Each filter creates a separate feature map, which subsequently get stacked and can be further processed by the following network layers. These filters improve during training. Maxpooling layers, on the other hand, are useful to achieve location-independence in the feature detection and reduce the dimensional load within the network. These layers forward only the highest value of small regions (e.g. 2x2) of the input map to the next layer, and in turn reduce the size of the feature map. The BatchNormalisation layer stabilizes and accelerates the training whereas the Dropout layers aid the network towards generalizing. All convolutional layers within a package feature the same kernel dimensions: The first convolutional package has kernels with the dimension of (7,7), decreasing for the next packages from (5,5) to two times (3,3) and ending for the last two packages at (2,2). 

The classification result for the two categories DSNB and NCQE is One-Hot encoded in a two-dimensional target vector. Due to the binary nature of the task, the binary cross-entropy function is chosen as the loss function for our network. The optimizer for this model is \textit{Adam} \cite{Adam}, which utilizes stochastic gradient decent based on adaptive estimations of first and second order moments. Nearly all convolutional and node layers use the \textit{ReLU} function as the activation function, only the final output layer applies \textit{softmax}. For the minibatch size we choose 128.

\paragraph{Network training.} We use three different class-balanced and shuffled sets: a training sample, a validation sample and a final test sample. The training sample consists of 185,000 class-balanced events with 92,500 positron events and 92,500 NCQE events. The validation sample contains  15,000 events and is used to monitor the progression of the training and indicate possible adjustments of hyperparameters. During the training, we assure that the CNN is not-overtrained by applying tensorflow callbacks such as \textit{Modelcheckpoint} and monitor the performance of the validation set for each epoch. Our final test sample is composed of 20,000 events between 10\,MeV and 30\,MeV (14,000 in the range 12--30\,MeV) and is used only once for the evaluation of the CNN performance. 

\subsection{Network performance} 
\label{sec:cnn_performance}

\begin{figure}[t!]
\centering
\includegraphics[width=0.8\textwidth]{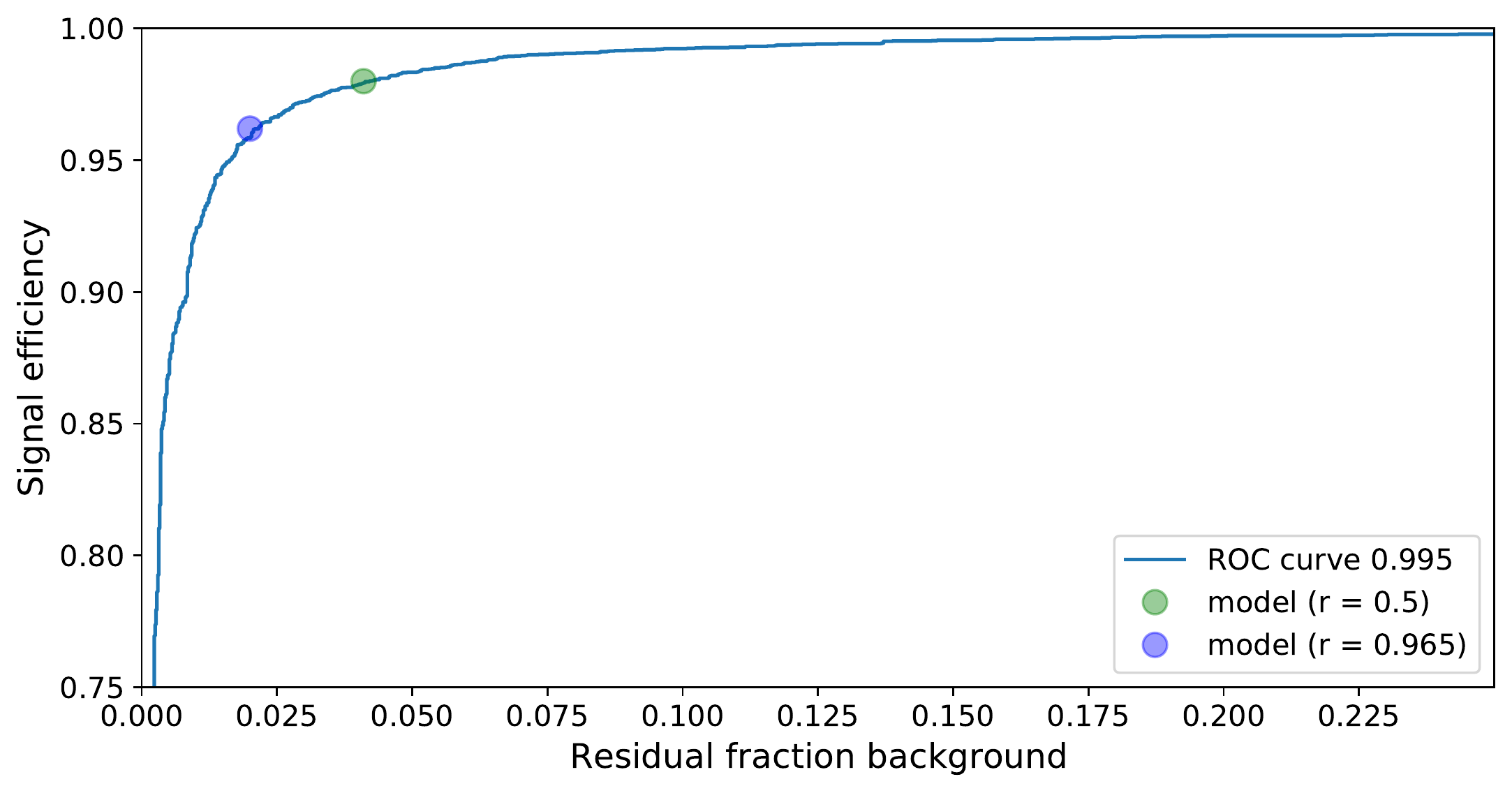}
\caption{ROC curve visualizing the CNN discrimination performance for the energy region 12-30 MeV. Depending on the discrimination threshold $r$, different combinations of signal efficiency and background rejection can be achieved. We mark the combination for the default value $r=0.5$ as well as a situation with enhanced background rejection for $r=0.965$ (cf.~Tab.~\ref{tab:Results}). }
\label{fig:ROC_results}
\end{figure}

As is customary, the CNN is assigning two continuous result parameters $r_i=[0..1]$ for all tested events. We are interested in the first result parameter of our network which corresponds to the likelihood for the event to be an DSNB event.  Based on the final test sample, we find a high accuracy for the classification for the relevant energy region 12-30 MeV. For a standard threshold value of $r=0.5$, DSNB (NCQE) events are correctly classified in 98\% (95.9\%) of all cases. This means that in case $r\geq0.5$ is applied as a selection criterion, 98\% of all DSNB but only 4.1\% of the NCQE will be selected.  
The CNN classification performance can be adapted by choosing a different threshold value $r$. This is exemplified by the Receiver Operating Characteristic (ROC) curve displayed in Fig.~\ref{fig:ROC_results} that displays possible combinations of signal efficiency and residual background fractions depending on the choice for the threshold $r$. 
For instance, $r= 0.965$ corresponds to a DSNB signal efficiency of 96.2\% and a NCQE background residual of only 2\% and hence provides better background rejection capabilities than the standard threshold of $r =0.5$. We choose $r=0.5$ and $r=0.965$ for our discussion in Sec.~\ref{sec:results}.




\begin{table}[!t]
\centering
\begin{tabular}{c|cc}
\toprule
threshold & DSNB signal & NCQE background \\
value & efficiency & residual  \\
\midrule
0.500 & 0.980 & 0.041 \\
0.965 & 0.962 & 0.020 \\
\bottomrule
\end{tabular}
\caption{Network performance for events with energies between 12-30 MeV in the test sample. The energy region 12-30 MeV is populated by $\sim$ 14,000 events and contains 56\% DSNB and 44\% NCQE events due to the underlying nature of the energy distribution. Both the DSNB signal efficiency and the NCQE background residual depend on the threshold value chosen for classification.} 
\label{tab:Results}
\end{table}

\subsection{Systematic effects} 
\label{sec:bias}

We have performed several tests of the CNN and its results to evaluate its performance, understand the main criteria for discrimination and to discover potential biases in the classification. 



\paragraph{Saliency maps.} A powerful tool to understand which features the CNN relies on for event classification are the saliency maps introduced in Sec.~\ref{sec:data_prep}. These maps are produced using guided back-propagation, i.e.~a heat map is created by following back only positive contributions (positive gradient information) of a signal on the output layer through all convolutional layers of the CNN to the input hit maps. By this back-propagation, the resulting saliency map ranks the contribution of each pixel in the hit maps according to a metric of importance for the CNN's classification decision \cite{Sailency} and thus permit inspection of the visual features the network is mostly relying on. 

\begin{figure}[!h]
\centering

\includegraphics[scale=0.38]{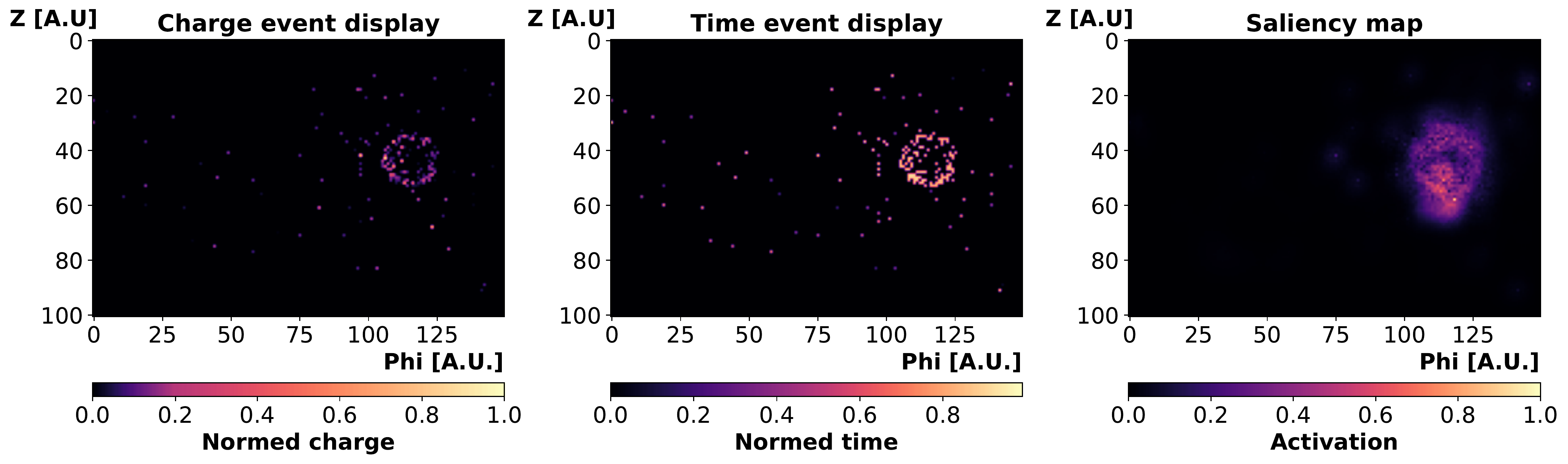}
\caption{From left to right, normalized charge and time hit maps and saliency map of a mis-classified NCQE event featuring a single strong ring. This class of events is more frequent at high visible energies (in this case: 26.5\,MeV).}
\label{fig:WrongClassification}
\end{figure}

\begin{figure}[!tbp]
  \centering
  \begin{minipage}[b]{0.67\textwidth}
    \includegraphics[width=\textwidth]{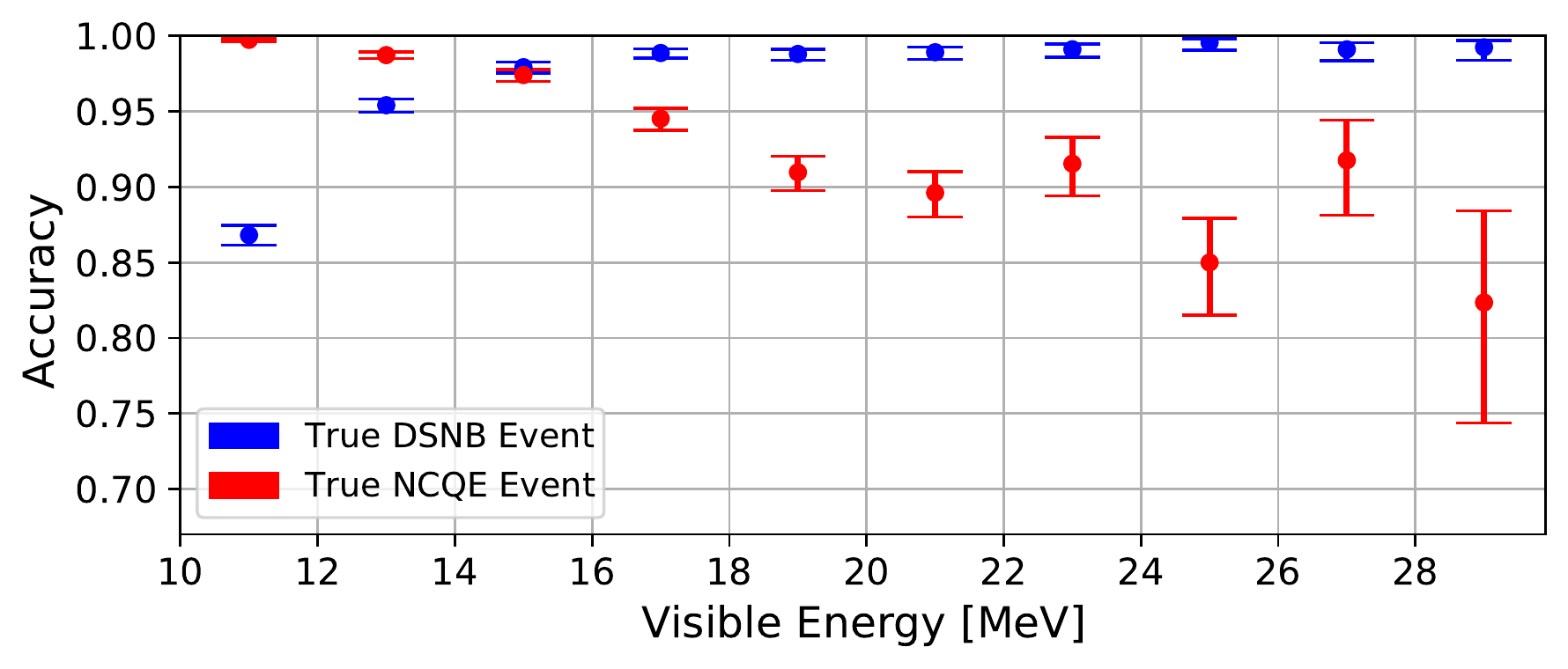}
    \caption{CNN classification accuracy as a function of visible energy for DSNB and NCQE events for r=0.5. The 1$\sigma$-error bars have been determined based on the Clopper-Pearson method. Despite the overall high accuracy, some energy dependence is apparent (see text). }
 \label{fig:class_energy}
  \end{minipage}
  \hfill
  \begin{minipage}[b]{0.3\textwidth}
    \includegraphics[width=\textwidth]{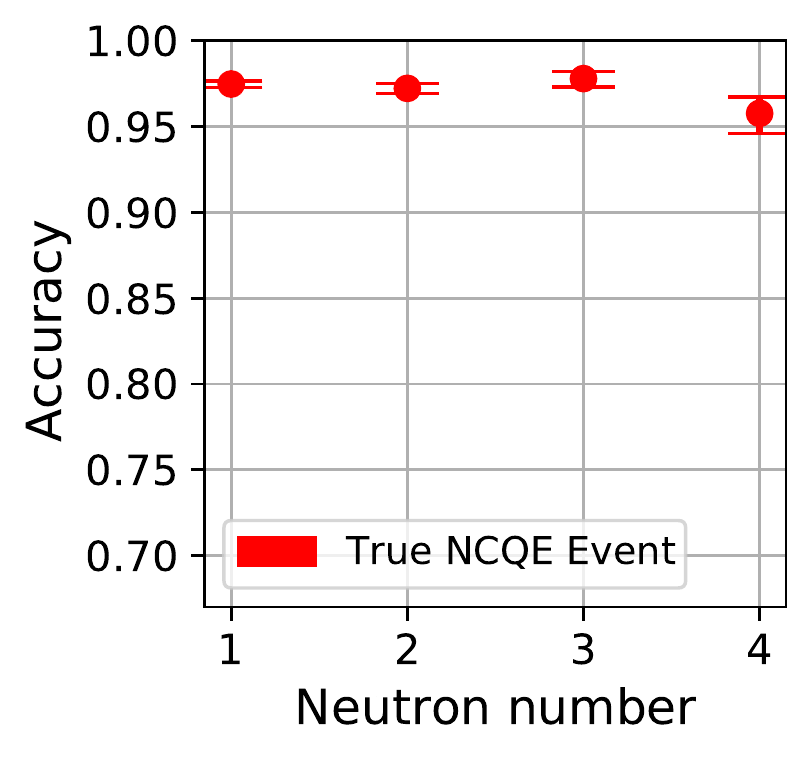}
     \caption{CNN accuracy as a function of the true neutron number of NCQE events. A weak dependence is visible.}
 \label{fig:class_neutron}
  \end{minipage}
  
\end{figure}

The two saliency maps we include with the representative DSNB and NCQE in Fig.~\ref{fig:event_displays} suggest that the CNN indeed reacts to the presence of (multiple) rings in the hit maps. We have visually checked the saliency maps of a selection of both correctly and mis-classified DNSB and NCQE events and find that -- as expected -- the main positive criterium for DSNB identification is the presence of a single strong Cherenkov ring. In accordance with this pattern, figure \ref{fig:WrongClassification} displays a NCQE event featuring a single Cherenkov ring that the CNN wrongly classifies as DSNB. 
\paragraph{Energy dependence.} As described in Sec.~\ref{sec:data_prep}, we are able to remove any first-order dependence of the event classification on its visible energy by normalizing the charge values registered by the PMTs for each hit map individually. This precaution has been introduced to prevent a classification of events based on the different spectral shapes. Nevertheless, we observe some energy-dependence in the classification accuracies displayed in Figure \ref{fig:class_energy}: while the accuracy is rising with visible energy for the DSNB, the inverse trend is visible for NCQE events. From visual checks of mis-classified NCQE events at visible energies above 26\,MeV, we deduce that NCQE events featuring a single strong Cherenkov ring are indeed more frequent at higher than at lower energies. Still, we find no evidence that the classification relies in this case on the spectral shape instead of topological features. Given the relative spectral ratios, such an effect would be expected to induce an opposite trend in classification efficiencies.

\paragraph{Neutron-number dependence.} It is conceivable that NCQE events might be featuring a difference in prompt event topologies depending on the number of knock-out neutrons $n$ present in the final state. As described in Sec.~\ref{sec:simulation}, we consider only events with a single {\it detected} neutron as background for the DSNB search. However, events with $n>1$ will contribute in case $(n-1)$ are not detected. We have studied a possible dependence of the CNN classification accuracy by regarding these event types individually. However, we find only a very mild $n$-dependence of the NCQE classification.

\paragraph{Topology bias.} CNNs often run the risk to pick up on localized features of the input images. In the present case, this might be caused by different average hit frequencies in the PMT hit maps for the DSNB and NCQE samples that might for instance be caused by non-uniformities in the event generation or geometries (both vertex position and direction). To exclude this, we first produce average maps for both charge and time by summing all events of one event category, i.e.~separately for both DSNB and NCQE event samples (70,000 maps each). The resulting maps are subtracted from each other and investigated for a local net difference. However, we find the fluctuations in these difference maps compatible with zero and can thus exclude such a bias. 

\paragraph{Channel preference.} Both the charge and the time hit maps represent individual input channels for the CNN. As laid out in Sec.~\ref{sec:cnn_arch}, they are combined to a common feature map by the first convolutional layer. To understand whether there is a preference of the network to time or charge data, we feed the trained CNN only one of the input hit maps (while providing an empty map in the other). While the full classification is preserved when only provided with time hit maps, the CNN performs significantly worse when offered only charge data.

However, it would be wrong to conclude from this result that the timing information is more relevant than the charge data. To test this hypothesis, we have trained two CNNs of identical structure but providing them in one case only with the time hit maps, in the other only with charge information. In fact, both networks are able to reach classification efficiencies very similar to that of our original two-channel CNN. Our conclusion is that the CNNs focus primarily on the spatial hit patterns (i.e.~hit vs.~empty PMTs), while the exact timing and charge information is of only little importance. This is relevant since it means that the performance of similar networks in other experiments will mostly depend on the granularity of the photo sensor array.

\section{Application to DSNB searches}
\label{sec:results}

\begin{figure}[!h]
  \centering
  \begin{minipage}[b]{0.5\textwidth}
    \includegraphics[width=\textwidth]{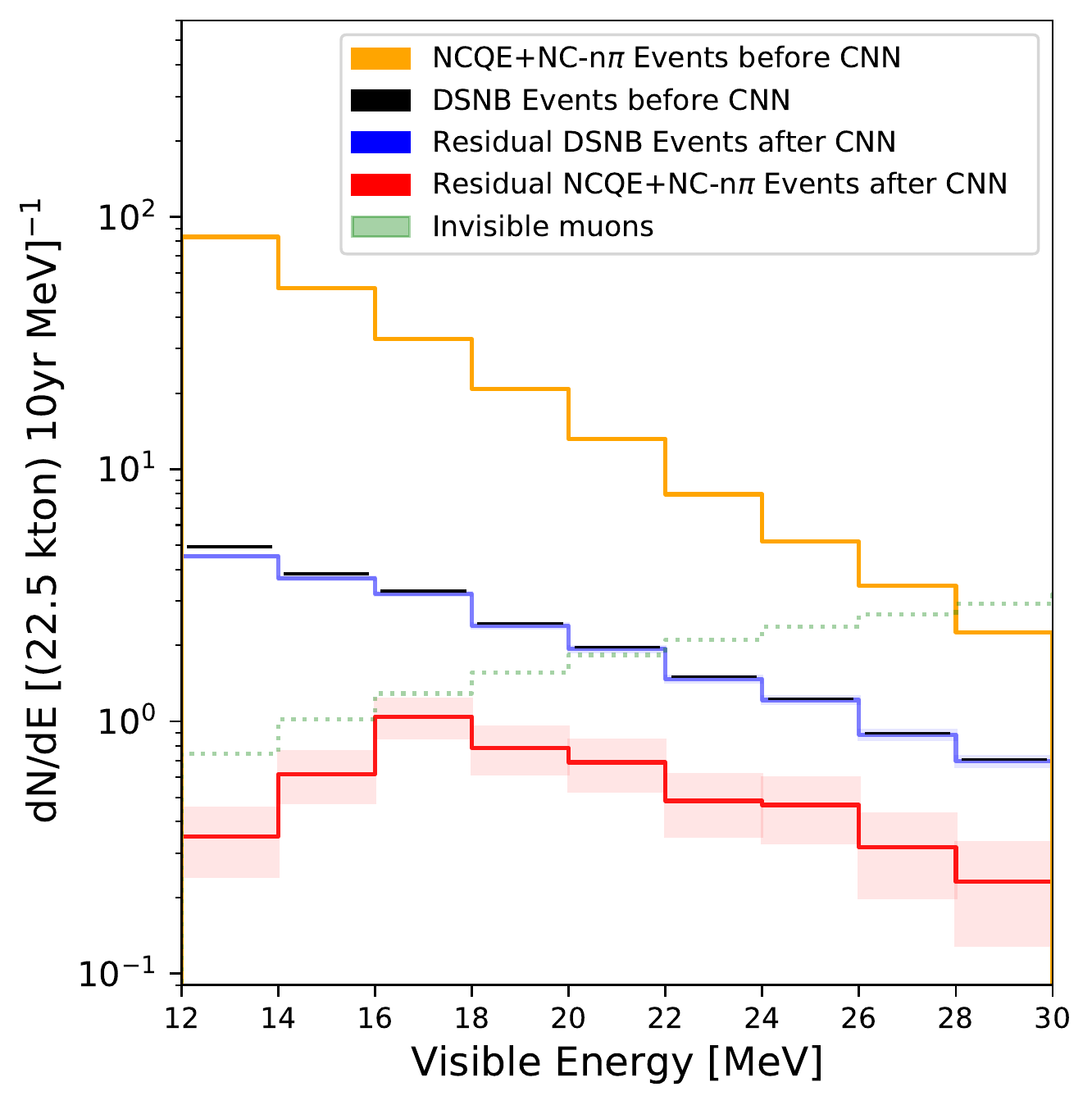}
  \end{minipage}
 \caption{The expected event rates for DSNB signal ({\it blue, black lines}) and NC background ({\it orange, red lines}) before and after the application of the CNN classification. We use the strict threshold value of $r\geq 0.965$ for the CNN (Sec.~\ref{sec:cnn_performance}). 
 For comparison, we show the spectral contributions from invisible muons ({\it green dotted line}) that won't be affected by CNN classification \cite{MuonRate}.}
 \label{fig:class_power}
\end{figure}

In a final step, we scale the statistics of our DSNB and NC test samples to the numbers of signal and background events expected for a measuring time of 10 years in our simplified SK-Gd-like detector. For the corresponding exposure of 225\,kt$\cdot$yrs, we expect 20 DSNB and 221 NC atmospheric events (148 NCQE plus 73 NC-$n\pi$ events) in the observation window. The corresponding visible-energy spectra are displayed in Fig.~\ref{fig:class_power}. Neglecting the invisible-muon background, the initial signal-to-background (S:B) ratio inside the observation window is about 1:11. Application of the CNN significantly improves the situation: While for the default discrimination threshold of $r\geq0.5$ almost no DSNB signal strength is lost (-2\%), the NCQE background is reduced to 4\% of its original rate. Hence, the S:B ratio improves to about 2:1. In case of a more stringent cut $r\geq0.965$, this ratio is further improved to 4:1.

\section{Conclusions and Outlook}
\label{sec:conclusions}

The upcoming measurements by SK-Gd and at a later point HK-Gd bear a great promise for the first detection of the elusive signal of the DSNB. Based on state-of-the-art event generators and existing measurements of this background in SK, we show in this article that NC interactions of atmospheric neutrinos pose a potentially significant background to this search. However, the event topologies of the prompt events are expected to be sufficiently different for DSNB and NCQE events such that a suppression of this background becomes feasible.

In the present paper, we have investigated a novel method to discriminate the two event categories not based on traditional event reconstruction parameters but using directly  the topology of the hit patterns recorded by the PMTs. For this, we have used a simplified detector layout and trained a CNN to distinguish signal and background events. The performance of the CNN is excellent, reaching 98\% signal efficiency and 96\% background rejection. After classification, the DSNB signal in the detection window (12--30\,MeV) exceeds the residual atmospheric background level at a ratio of 4:1. In this, the discrimination power relies mostly on the inherent differences in event topologies (single strong vs.~multiple washed-out rings) and thus primarily depends on the granularity of the light readout (i.e.~the number of PMTs per photosensitive area).

CNN-based classification thus provides a real and systematically largely independent alternative to conventional methods of event selection for DSNB searches in large water-Gd detectors. 


\acknowledgments

We thank Julia Sawatzki (TU Munich), Jan Peter Lommler and Nils Brast (both JGU Mainz), Mark Vagins (IPMU, Tokyo University and UC Irvine), and Koun Choi (IBS, Daejeon) for the valuable discussions and comments to this manuscript.

\bibliographystyle{JHEP}
\typeout{}
\bibliography{main}

\end{document}